\documentclass[fleqn,twoside]{article}
\usepackage{espcrc2}

\usepackage{graphicx}

\usepackage[figuresright]{rotating}

\newcommand{\AmS}{{\protect\the\textfont2
  A\kern-.1667em\lower.5ex\hbox{M}\kern-.125emS}}

\title{Decays $\tau^{-} \rightarrow \eta(\eta^\prime)\pi^{-}\pi^{0}\nu_{\tau}$
and CVC}
\author{ V. Cherepanov\address[MCSD]
{III. Physikalisches Institut B, 
RWTH Aachen University, D-52056 Aachen, Germany }, % 
  S. Eidelman\address{Budker Institute of Nuclear Physics, 630090, 
Novosibirsk, Russia \\ and Novosibirsk State University, 630090, Novosibirsk,
Russia}\thanks{Speaker}}
\begin{document}

\begin{abstract}
We use experimental data on 
$e^{+}e^{-} \rightarrow \eta\pi^{+}\pi^{-}$ and
$\tau^{-} \to \eta\pi^{-}\pi^{0}\nu_{\tau}$ to test 
conservation of vector current (CVC) by comparing the predicted
hadronic spectrum and branching fraction with the $\tau$ decay data. 
Based on the corresponding $e^+e^-$ data and CVC, we also calculate
the branching fraction of 
$\tau^{-} \to \eta^\prime\pi^{-}\pi^{0}\nu_{\tau}$ decay. 
\vspace{1pc}

\end{abstract}

% typeset front matter (including abstract)

\maketitle

\section{Introduction}
Low energy $e^{+}e^{-}$ annihilation into hadrons is
a source of valuable information about the interactions of
light quarks. Precise measurements of the exclusive cross
section appear important for different applications like, 
e.g., determination of various QCD parameters --- quark
masses, quark and gluon condenstates~\cite{shifman}, calculation of 
the hadronic contributions to the muon anomalous magnetic moment
and running fine structure constant~\cite{eidelman1}. 

The hypothesis of conserved vector current and isospin symmetry
relate to each other the isovector part of $e^{+}e^{-} \rightarrow $ 
hadrons and corresponding (vector current) hadronic decay of the 
$\tau$ lepton~\cite{tsai,thacker}. This follows from the deep relation 
between weak and electromagnetic (EM) interactions. 
%If one neglects quark masses, QCD is invariant
%under a transformation replacing quark flavours.
The vector weak current and the isovector part of the electromagnetic vector 
current are different components of the same vector current, so that the
matrix element of these currents 
%$\bar{u}\tau^+\gamma_{\alpha}d$ and  
%$\bar{u}\gamma_{\alpha}u-\bar{d}\gamma_{\alpha}d$ 
must be identical assuming SU(2) symmetry. In this case the weak 
isovector current is assumed to be conserved in analogy
to the EM current. This assumption is the CVC hypothesis.

As a consequence, hadronic currents describing vector $\tau$ decays 
and low energy (up to the $\tau$ lepton mass)
$e^{+}e^{-}$ annihilation are related and can be obtained one from another.
These relations allow one to use an independent high-statistics data sample 
from $\tau$ decays for increasing the accuracy of the prediction of 
the spectral functions directly measured in $e^+e^-$ 
annihilation~\cite{alemany}.
 
The very first application of this idea was very fruitful~\cite{alemany}, 
but increasing experimental precision in both $e^{+}e^{-}$ 
and $\tau$ sectors revealed unexpected problems: the $2\pi$ and
$4\pi$ spectral functions from $\tau$ decays were significantly
higher than those obtained from  $e^{+}e^{-}$~\cite{davier1,davier2}.
Although tension in this sector has been recently somewhat decreased
after a reestimation of the isospin breaking corrections~\cite{davier3},  
it is important to understand the reasons causing the
deviations between the spectral functions. One of the necessary steps 
is to perform a systematic test of CVC relations using all available 
experimental information on various final states. 

For the vector part of the weak hadronic current, the distribution of 
the mass of the produced hadronic system is:
\begin{equation}
\frac{d\Gamma}{dq^{2}} = \frac{G_{F}|V_{ud}|^{2}S_{EW}}{32\pi^{2}m^{3}_{\tau}}
(m^{2}_{\tau}-q^{2})^{2}(m^{2}_{\tau}+2q^{2})v(q^{2}),
\end{equation}

\noindent
where the spectral function $v(q^{2})$ is given by the expression:
\begin{equation}
v(q^{2})=\frac{q^{2}\sigma^{I=1}_{e^{+}e^{-}}(q^{2})}{4\pi^{2}\alpha^{2}},
\end{equation}
and $S_{EW}$ is an electroweak correction equal to 1.0194~\cite{marciano}.

Since the vector part of the weak current has a positive $G$-parity, the
allowed quantum numbers for hadronic decays are:
\begin{equation}
J^{PG}=1^{-+}, \tau \rightarrow 2n\pi\nu_{\tau},\omega\pi\nu_{\tau},
\eta\pi\pi\nu_{\tau},...
\end{equation} 
   
After integration the branching fraction of the $\tau$ decay is
%\large{
$\frac{{\cal B}(\tau^{-} \rightarrow X^{-}\nu_{\tau})}
{{\cal B}(\tau^{-} \rightarrow e^{-}\nu_{e}\nu_{\tau})}=
\frac{3|V_{ud}|^{2}S_{EW}}{2\pi\alpha^{2}}\times$
\begin{equation}
\int_{4m^{2}_{\pi}}^{m^{2}_{\tau}}dq^{2}\frac{q^{2}}{m^{2}_{\tau}}
(1-\frac{q^{2}}{m^{2}_{\tau}})
(1+2\frac{q^{2}}{m^{2}_{\tau}})\sigma_{e^{+}e^{-}}^{I=1}(q^{2})
\end{equation}

In this work we focus on two specific 
final states of $e^+e^-$ annihilation ($\tau^-$ decay): 
$\eta\pi^+\pi^-$ ($\eta\pi^-\pi^0\nu_{\tau}$)  and 
$\eta^\prime\pi^+\pi^-$ ($\eta^\prime\pi^-\pi^0\nu_{\tau}$).
Theoretical calculations for the corresponding decay modes of the $\tau$ 
lepton based on CVC were earlier performed by many authors, 
see the bibliography in Ref.~\cite{eidelman2,cherepanov}.
New comparison of CVC-based predictions with measurements of $\tau$ lepton 
decays were motivated by recent progress of experiments on $\tau$ decays  
as well as by the updated information from $e^{+}e^{-}$ annihilation 
into hadrons, coming mostly from the BaBar~\cite{aubert} and 
SND~\cite{achasov} collaborations. 

For numerical estimates we use the value of the electronic 
branching ${\cal B}(\tau \rightarrow e\bar{\nu_{e}}\nu_{\tau})$ = 
(17.85 $\pm$ 0.05)\% and $|V_{ud}|^{2}=0.97425 \pm 0.00022$ recommended by 
RPP-2010~\cite{nakamura}.

 \section{$\tau^{-} \rightarrow \eta\pi^{-}\pi^{0}\nu_{\tau}$}
The reaction $e^{+}e^{-} \rightarrow \eta\prime\pi^{+}\pi^{-}$ was 
recently studied by the BaBar collaboration using ISR in the broad energy 
range from 1~GeV to 3~GeV~\cite{aubert} and by the SND collaboration
in the energy range from 1.1~GeV to 1.4~GeV~\cite{achasov}. 
Earlier measurements were performed at the ND~\cite{druzhinin}, 
CMD-2~\cite{akhmetshin} detectors from 1.25 to 
1.4~GeV and at the DM1~\cite{delcourt} and DM2~\cite{antonelli} 
detectors from 1.4 to 2~GeV. The results of various measurements 
are shown in Fig.~1. In general, they are in fair agreement 
with each  other within errors although below 1.4 GeV the values of the 
cross section from BaBar are somewhat higher than those of the previous 
experiments. Above this energy, the results from BaBar are higher than 
those from DM2, whereas  they are in good agreement with much less precise
data of DM1. A more detailed information about the  data samples used 
can be found in Table~\ref{table1}.
\begin{center}
\begin{table*}[htb]
\caption{Summary of $e^{+}e^{-} \rightarrow \eta\pi^{+}\pi^{-}$ data}
\label{table1}
\newcommand{\m}{\hphantom{$-$}}
\newcommand{\cc}[1]{\multicolumn{1}{c}{#1}}
\renewcommand{\tabcolsep}{2pc} % enlarge column spacing\renewcommand{\arraystretch}{1.2} % enlarge line spacing
\begin{tabular}{@{}lllll}
\hline
Group & $\sqrt{s}$, GeV  & N$_{\rm points}$ & $\Delta_{\rm stat}$, \% & 
$\Delta_{\rm syst}$, \%  \\
\hline
ND, 1986            & 1.25 - 1.40   & 3   & 50 - 100 & 10 \\
CMD-2, 2000         & 1.25 - 1.40   & 6   & 30 - 60  & 15 \\
SND, 2010           & 1.17 - 1.38   & 6   & 15 - 60  & 10.5 \\
DM1, 1982           & 1.40 - 1.80   & 4   & 30 - 60  & 10 \\
DM2, 1988           & 1.35 - 1.80   & 10  & 25 - 60  & 10\\
\hline
BaBar, 2007         & 1.00 - 1.80   & 16  & 10 - 60  & 8 \\
\hline
\end{tabular}\\[2pt]
\end{table*}
\end{center}
We calculated the branching fraction of 
$\tau^{-} \rightarrow  \eta\pi^{-}\pi^{0}\nu_{\tau}$ 
decay expected from the above mentioned $e^{+}e^{-}$ data using the 
relation (4). 
The direct integration of experimental points in the energy range 
from 1.25~GeV to the $\tau$ mass using older data samples gives 
for the branching fraction (0.130 $\pm$ 0.015)\%
in agreement with the previous estimate~\cite{eidelman2}, while that 
based on the BaBar data gives
(0.165 $\pm$ 0.015 )\%, where we took into account the 8\% systematic error 
claimed by the authors~\cite{aubert}. Since two results differ by more 
than one standard deviation, we follow the PDG prescription and 
inflate the uncertainty of their weighted average by a scale factor of 1.67.
This gives for the CVC-based branching fraction (0.147 $\pm$ 0.018)\% 
in the energy range (1.25 - 1.77)~GeV. Finally,
we add the contribution of the low energy range from 1.0~GeV to 1.25~GeV 
(based on the BaBar data set)
to obtain the total CVC expectation of (0.153 $\pm$ 0.018)\%. It can be 
compared to the measured branching fractions which are shown in Table 2 
and include both older results from CLEO~\cite{artuso} and 
ALEPH~\cite{busculic}
and the recent experimental result from Belle~\cite{inami}. 
Our estimate is consistent within errors with all $\tau$ measurements 
as well as with their average of (0.139 $\pm$ 0.008)\%, which is 
0.9$\sigma$ lower than our CVC-based prediction.

In addition, we show in Fig.~2  the spectrum of  $\eta\pi^{-}\pi^{0}$
masses obtained by Belle and in Fig.~3 compare it (after subtracting 
the background) with the spectral function calculated from all available 
$e^{+}e^{-}$ data using relation (1). In general, the two spectral 
functions are in fair agreement with each other except a few points
near the lower and higher boundaries, where 
the values of $F^{2}_{\eta\pi\pi}[ee]$  are close to zero.

\begin{figure}[htb]

\vspace{5pt}
\includegraphics[scale=0.4]{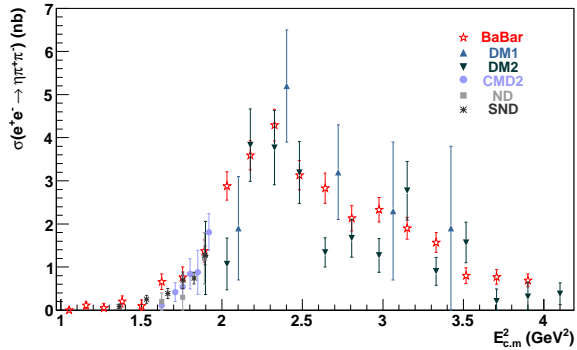}
%\includegraphics[scale=0.4]{cross.eps}
%\framebox[55mm]{\rule[-21mm]{0mm}{43mm}}

\caption{Cross section of the process $e^+e^- \rightarrow \eta\pi^+\pi^{-}$.}

%\label{fig:largenenough}

\end{figure}

\begin{figure}[htb]

\vspace{7pt}
\includegraphics[scale=0.4]{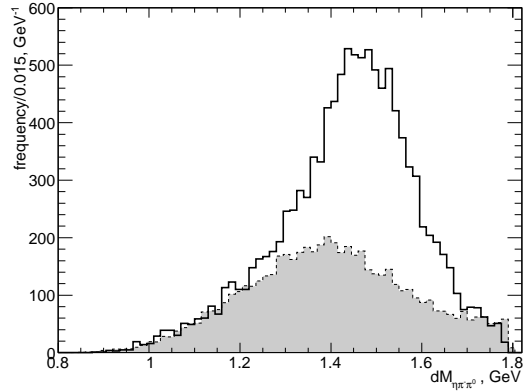}
%\includegraphics[scale=0.4]{Distrib_Inami_advanced.eps}
%\framebox[55mm]{\rule[-21mm]{0mm}{43mm}}

\caption{The mass spectrum of $\tau$ decay to $\eta\pi\pi\nu_{\tau}$ 
obtained by the Belle collaboration, the shaded histogram shows a background}

%\label{fig:largenenough}

\end{figure}

\begin{figure}[htb]

\vspace{7pt}
\includegraphics[scale=0.4]{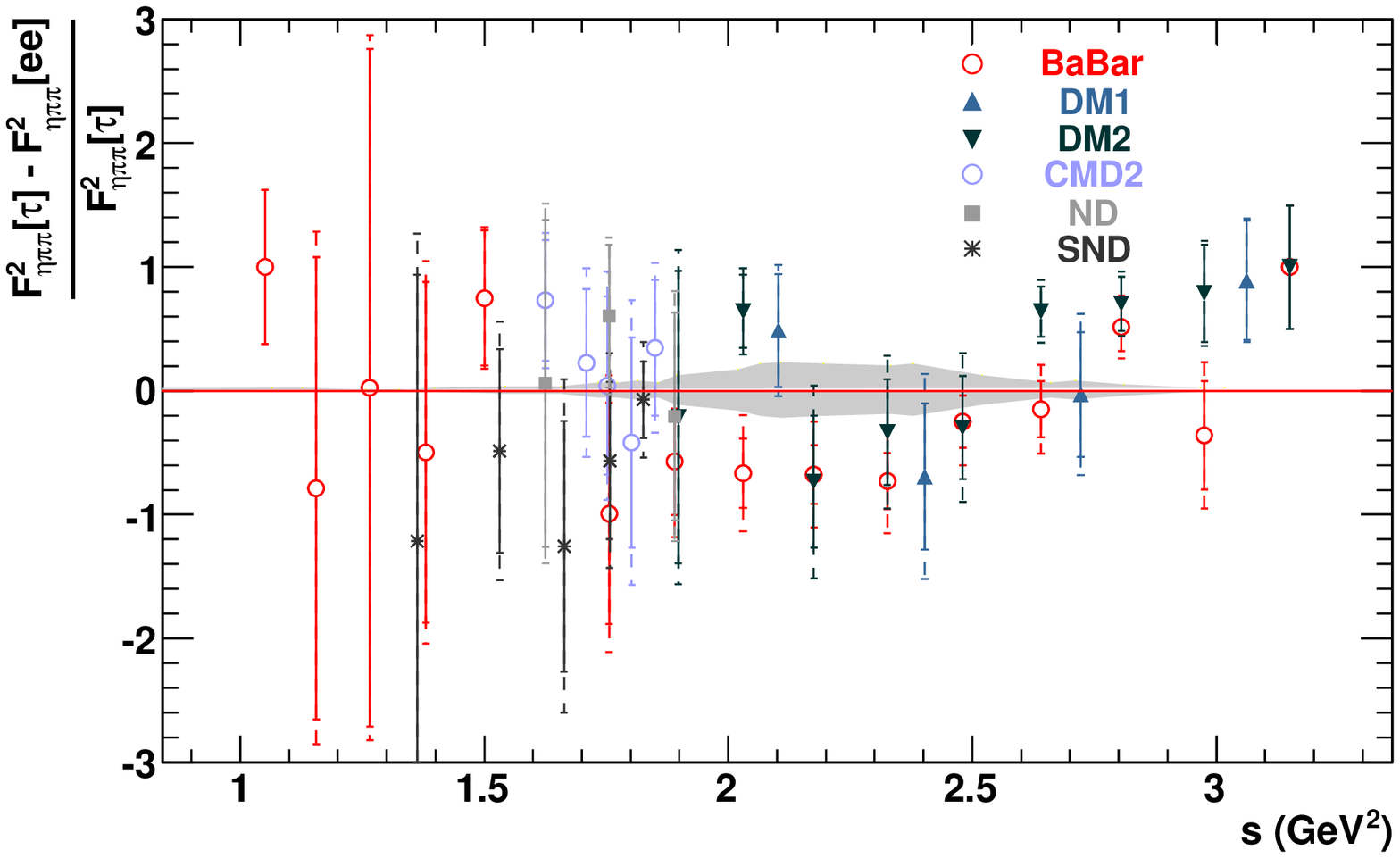}
%\includegraphics[scale=0.4]{spectrCompsl-slide12.eps}
%\framebox[55mm]{\rule[-21mm]{0mm}{43mm}}

\caption{Comparison of the $\tau$ and $e^+e^-$ spectral functions 
taking into account a  5.3\% syst. error of Belle }

%\label{fig:largenenough}

\end{figure}

%The average of the experimental results gives 
%$\cal B(\tau^{-} \rightarrow \eta\pi^{-}\pi^{0}\nu_{\tau})$
% = (0.139 $\pm$ 0.008)\%, which is 0.9$\sigma$ lower than our 
%CVC-based prediction.

It is also interesting to compare our result wuth earlier theoretical 
estimates of this branching fraction, see Table 2. It can be seen that 
the older predictions based on the $e^{+}e^{-}$  data and CVC agree with 
the much more accurate result of this work,
which uses more precise data, in particular, the recent data sample 
of BaBar. Other predictions,
which are more theoretically driven and use low-energy effective Lagrangians, 
show a much larger spread of the results.

\begin{table*}[htb]
\begin{center}
\caption{Experimental values of 
${\cal B}(\tau^{-} \rightarrow \eta\pi^{-}\pi^{0}\nu_{\tau})$}
\label{table:2}
\newcommand{\m}{\hphantom{$-$}}
\newcommand{\cc}[1]{\multicolumn{1}{c}{#1}}
\renewcommand{\tabcolsep}{2pc} % enlarge column spacing
\renewcommand{\arraystretch}{1.2} % enlarge line spacing
\begin{tabular}{@{}llllll}
\hline
Group       &    & ${\cal B}$,\% &  & Ref. & \\
\hline
CLEO, 1992  &         & 0.170 $\pm$ 0.020 $\pm$ 0.020  &   &\cite{artuso}  &\\
ALEPH, 1997 &         & 0.180 $\pm$ 0.040 $\pm$ 0.020 &   &\cite{busculic}  & \\
Belle, 2009 &         & 0.135 $\pm$ 0.003 $\pm$ 0.007&    &\cite{inami}   & \\
\hline
\end{tabular}\\[2pt]
\end{center}
\end{table*}

\begin{table*}[htb]
\begin{center}

\caption{Theoretical predictions for  
${\cal B}(\tau^{-} \rightarrow \eta\pi^{-}\pi^{0}\nu_{\tau})$}

\label{table:3}
\newcommand{\m}{\hphantom{$-$}}
\newcommand{\cc}[1]{\multicolumn{1}{c}{#1}}
\renewcommand{\tabcolsep}{2pc} % enlarge column spacing
\renewcommand{\arraystretch}{1.2} % enlarge line spacing
\begin{tabular}{@{}lllllll}
\hline
Method       &    & ${\cal B}$,\% & & & Ref. &     \\
\hline
$\rho\prime$      &          & $\sim$0.3     &  &  & \cite{pich}   & \\
CVC               &          & $\sim$0.15    &  &  & \cite{gilman}  & \\
Eff.~Lagr.         &          & 0.14$^{+0.19}_{-0.10}$ & &  &\cite{braaten} &   \\
Eff.~Lagr.         &          & 0.18-0.88      & &  & \cite{kramer}  & \\
CVC               &          & 0.13$\pm$0.02 &  &  & \cite{eidelman2} &  \\
CVC               &          & 0.14$\pm$0.05  & &  & \cite{narison} &  \\
CVC + Eff.~Lagr.   &          & $\sim$ 0.19    & &  & \cite{decker}  & \\
Eff.~Lagr.         &          &  $\sim$ 0.1   &  &  & \cite{li}  & \\
\hline
\end{tabular}\\[2pt]
\end{center}
\end{table*}

\subsection{Cross section approximation}
For future applications of our results aimed at the improvement of the
existing Monte Carlo generators of $e^+e^-$ annihilation~\cite{PHOKHARA} 
and $\tau$ decay~\cite{TAUOLA} we perform the approximation of the 
cross section of the 
process  $e^{+}e^{-} \rightarrow \eta\pi^{+}\pi^{-}$. Its energy dependence 
is described by a sum of the 
$\rho\prime$(1450) and $\rho\prime\prime$(1700) contributions
(independently we check that the one from the $\rho(770)$ tail is 
negligible in this energy range):
\begin{equation}
\sigma_{\eta\pi\pi}(s) = \frac{F_{\eta\pi\pi}(s)}{s}|A_{\rho\prime} + 
A_{\rho\prime\prime}e^{i\delta_{\rho\prime - \rho\prime\prime}} |^{2},
\end{equation}
\begin{equation}
A_{V} = \frac{m^{2}_{V}\Gamma_{V}\sqrt{\sigma_{V}/F_{\eta\pi\pi}(m^{2}_{V})}}
{s-m^{2}_{V} + i\sqrt{s}\Gamma_{V}(s)},V=\rho\prime,\rho\prime\prime,
\end{equation}
where $m_{V}, \Gamma_{V},\sigma_{V}$ are mass, width and peak cross 
section of the intermediate $\rho\prime$ or $\rho\prime\prime$, 
$\delta_{\rho\prime - \rho\prime\prime}$ is the
relative phase of the $\rho\prime - \rho\prime\prime$ interference, 
and $F_{\eta\pi\pi}(s)$ is a smooth function arising from the decay
matrix element squared and the phase space, which is written as 
an integral over the kinematically allowed region
on the $E_{+}$--$E_{-}$ plane:

\begin{equation}
F_{\eta\pi\pi}(s)=\int\int dE_{-}dE_{+}|\vec{p}_{+}\times\vec{p}_{-}|^{2}R_{2\pi}(E_{+}E_{-}),
\end{equation}
where $E_{+},E_{+},\vec{p}_{+},\vec{p}_{-} $ are energies and momenta of 
pions, $|\vec{p}_{+}\times\vec{p}_{-}|^{2}$ is a factor
reflecting the properties of vector particle decay into three pseudoscalars 
and $R_{2\pi}(E_{+}E_{-})$ is a function characterizing 
$\rho$(770)$\rightarrow \pi^{+}\pi^{-}$ decay dynamics and can be written as:

\begin{equation}
R_{2\pi}(E_{+}E_{-})  = \frac{1}{ \frac{Q^{2}}{M^{2}_{\rho}} - 1 + i\frac{\sqrt{Q^{2}}\Gamma_{\rho}(Q^{2})}{M^{2}_{\rho}}  }
\end{equation}

Since our knowledge of decay channels of both $\rho\prime$(1450) and 
$\rho\prime\prime$(1700) mesons is rather poor, the energy-dependent widths
are written as: 

\begin{equation}
\Gamma(s) = \Gamma_{0}\frac{s}{M^{2}}
\end{equation}
where $\Gamma_{0}$ and $M$ are full width and mass of the resonances.

In view of the observed excess of BaBar cross sections over all others,
we perform two independent fits -- that of BaBar and all other data.
Results of the fits are shown in Fig.~4 together with the difference between 
them (the dashed line). As one can see, the difference is higher than
zero  in the whole energy range which may point at the fact
that cross sections obtained by BaBar are systematically somewhat higher 
than those from previous experiments. 

\begin{figure}[htb]

\vspace{9pt}
\includegraphics[scale=0.4]{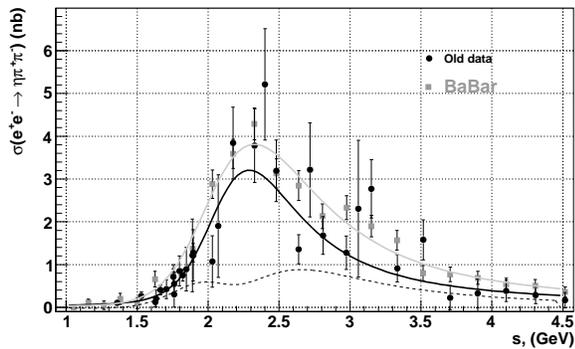}
%\includegraphics[scale=0.4]{crossFit.eps}
%\framebox[55mm]{\rule[-21mm]{0mm}{43mm}}

\caption{Cross section with optimal curves for two data samples }

%\label{fig:largenenough}

\end{figure}

 \section{$\tau^{-} \rightarrow \eta^\prime\pi^{-}\pi^{0}\nu_{\tau}$}
Recently the BaBar collaboration has presented the very first measurement
of the cross section of the process 
$e^+e^- \rightarrow \eta^\prime\pi^+\pi^-$~\cite{aubert}, see Fig.~5. 
The cross section of the process clearly shows resonance behavior with
a maximum slightly above 2 GeV. We fit the cross section 
assuming that it is described by a single resonance and 
parameterizing it with  the Breit-Wigner amplitude for production of 
three pseudoscalar mesons~\cite{akhmetshin2}. The following resonance 
parameters (mass, width and cross section at the peak) have been obtained:
\begin{figure}[htb]

\vspace{9pt}
\includegraphics[scale=0.4]{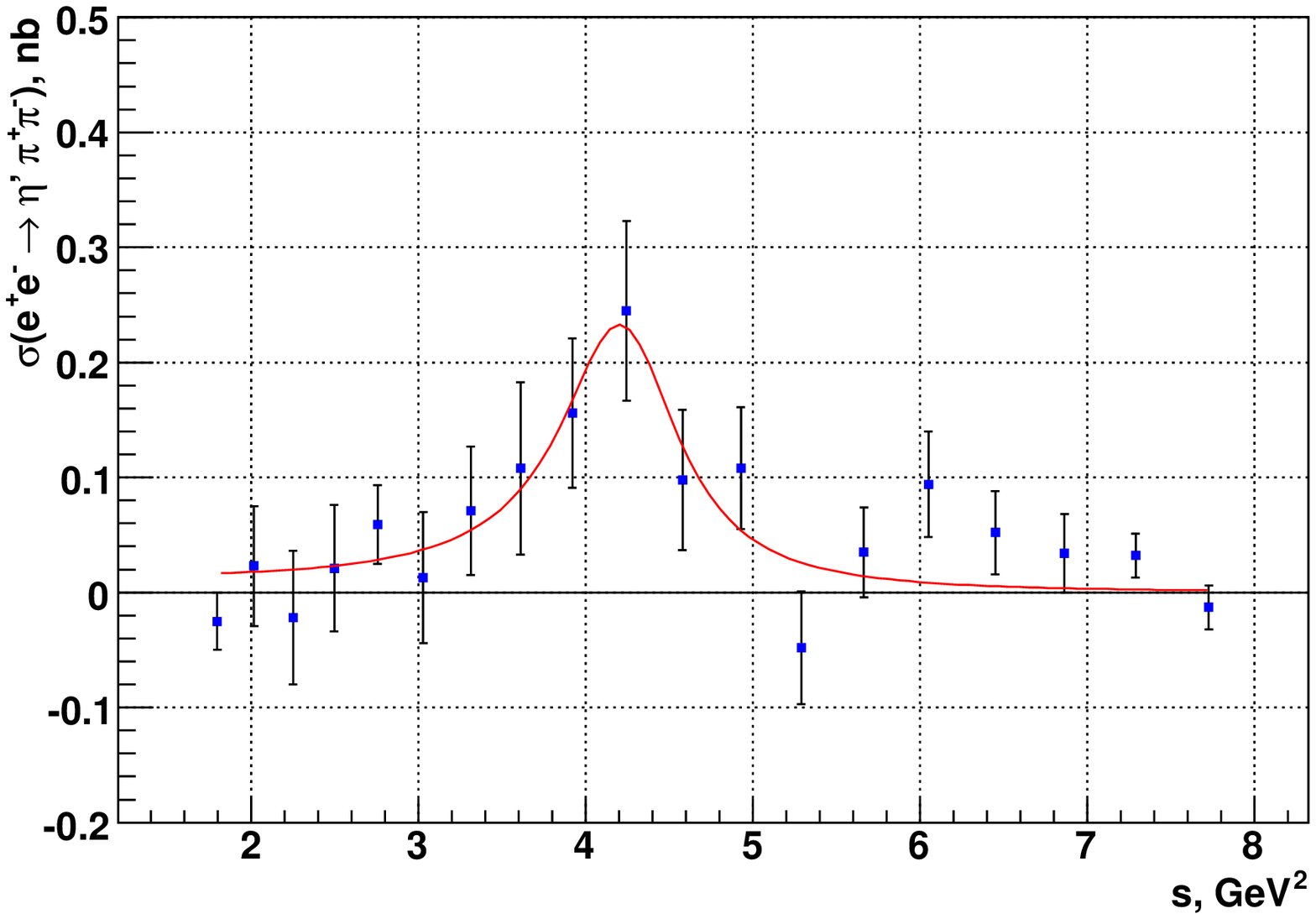}
%\includegraphics[scale=0.4]{vol4.eps}
%\framebox[55mm]{\rule[-21mm]{0mm}{43mm}}

\caption{Cross section of the process 
$e^+e^- \rightarrow \eta^\prime\pi^+\pi^{-}$.}
%$e^+e^- \rightarrow \eta\prime\pi^+\pi^{-}$.}
%\label{fig:largenenough}

\end{figure}

\begin{equation}
M=(2071 \pm 32) \ \ {\rm MeV},
\end{equation}
\begin{equation}
\Gamma=(214 \pm 76) \ \ {\rm MeV},
\end{equation}
\begin{equation}
\sigma_{0} = (0.223 \pm 0.073 \pm 0.022) \ \ {\rm nb}.
\end{equation}
Here the systematic error of $\sigma_{0}$ is 10\%  following the estimate
of the overall systematic error of the cross section in Ref.~\cite{aubert}.
Taking into account that we do not estimate systematic uncertainties
on mass and width, one can conclude that the resonance parameters are 
compatible with those of the $\rho$(2150)~\cite{nakamura}. 
Since the values of the cross section are very low and have large 
uncertainties, we do not integrate directly the experimental points
to estimate the branching fraction from CVC.
Instead, we use (4) and integrate the optimal curve
for the cross section up to the $\tau$ mass and obtain:
\small
\begin{equation}
\tiny {\cal B}(\tau^- \rightarrow \eta^\prime\pi^-\pi^0\nu_{\tau}) = 
(13.4 \pm 9.4 \pm 1.3 \pm 6.1)\times 10^{-6},
\end{equation}
\normalsize
where the first error is statistical (that of the fit), the second one 
is experimental systematic and the third is the model one estimated by using
the world average values of the $\rho$(2150) mass and width and varying them
within the errors. The obtained result is consistent with zero and we place
the following upper limit at 90\% CL using the method of Ref.~\cite{feldman}:
\begin{equation}
{\cal B}(\tau^{-} \rightarrow \eta^\prime\pi^{-}\pi^{0}\nu_{\tau})
<3.2 \times 10^{-5},
\end{equation} 
which is two and a half times more restrictive than the upper limit based on
the only existing measurement from CLEO~\cite{bergfeld}:
\begin{equation}
{\cal B}(\tau^{-} \rightarrow \eta^\prime\pi^{-}\pi^{0}\nu_{\tau})
<8 \times 10^{-5},
\end{equation}
but still an order of magnitude higher than a theoretical prediction 
${\cal B}(\eta^\prime\pi^{-}\pi^{0}\nu_{\tau})~\approx~4.4 \times 10^{-6}$ 
based on the chiral Lagrangian \cite{li}.

\section{Conclusion}
Using data on the processes 
$e^{+}e^{-} \to \eta(\eta^{\prime})\pi^-\pi^0\nu_{\tau}$ and CVC we obtained 
the following results:
\begin{itemize}
\item for $\tau^- \to \eta\pi^-\pi^0\nu_{\tau}$ the expected branching
fraction is (0.153 $\pm$ 0.018)\% compatible with the
world average of (0.139 $\pm$ 0.008)\%;
\item
the spectral functions of the $\eta\pi\pi$ system in $e^+e^-$ annihilation 
and $\tau$ decay are consistent;
\item for $\tau^- \to \eta^\prime\pi^-\pi^0\nu_{\tau}$ 
the upper limit on the branching fraction is $<3.2 \times 10^{-5}$ 
or 2.5 times smaller than
the experimental one of $<$ 8 $\times$ 10$^{-5}$, both at 90\% CL.
\end{itemize}

\section{Acknowledgments}
The authors are grateful to George Lafferty and his colleagues for the
excellent organization of the Workshop. We thank Henryk Czy\.{z},
Kenji Inami and Zbigniew W\c{a}s for useful
discussions. SE acknowledges the support
of the Organizing Committee. The participation of SE was partially supported
by the grants RFBR 10-02-08473 and 10-02-00695.


\begin{thebibliography}{9}

\bibitem{shifman} M. Shifman, A. Vainshtein, V. Zakharov, 
Nucl. Phys. B {\bf 147} (1979) 448.
\bibitem{eidelman1} S. Eidelman and F. Jegerlehner, Z. Phys. C {\bf 67} 
(1995) 585.
%\bibitem{cherepanov} V. A. Cherepanov and S. I.  Eidelman, 
%JETP Lett. {\bf 89} (2009) 429 [Pisma v ZhETF {\bf 89} (2009) 515].
\bibitem{tsai} Y. S. Tsai, Phys. Rev. D {\bf 4} (1971) 2821.
\bibitem{thacker}  H. B. Thacker and J. J. Sakurai, 
Phys. Lett. B {\bf 36} (1971) 103.
\bibitem{alemany} R. Alemany, M. Davier and A. H\"ocker, 
Eur. Phys. J. C  {\bf 2} (1998) 123.
\bibitem{davier1} M. Davier  et al., Eur. Phys. J. C {\bf 27} (2003) 497.
\bibitem{davier2} M. Davier  et al., Eur. Phys. J. C {\bf 31} (2003) 503.
\bibitem{davier3} M. Davier  et al., Eur. Phys. J. C {\bf 68} (2010) 127.
\bibitem{marciano} W. J. Marciano and A. Sirlin,  Phys. Rev. Lett. 
{\bf 61} (1988) 1815.
\bibitem{eidelman2} S. I. Eidelman and V. N. Ivanchenko,  
Phys. Lett. B {\bf 257} (1991) 437.
\bibitem{cherepanov} V. A. Cherepanov and S. I.  Eidelman, 
JETP Lett. {\bf 89} (2009) 429 [Pisma v ZhETF {\bf 89} (2009) 515].
%\bibitem{nakamura}K. Nakamura et al. [Particle Data Group], 
%J. Phys. G {\bf 37} (2010) 075021.
%\bibitem{amsler} C. Amsler et al.,  Phys. Rev. Lett. B {\bf 667} (2008) 1.
\bibitem{aubert} B. Aubert et al.,  Phys. Rev. D {\bf 76} (2007) 092005,\\
Erratum-ibid, D {\bf 77} (2008) 119903.
\bibitem{achasov} M. N. Achasov et al,. JETP Lett. {\bf 92} (2010) 80  
[Pisma v ZhETF {\bf 92} (2010) 84].
\bibitem{nakamura} K. Nakamura et al. [Particle Data Group], 
J. Phys. G {\bf 37} (2010) 075021.
\bibitem{druzhinin} V. P. Druzhinin et al.,  Phys. Lett. B {\bf 174} (1986) 115.
\bibitem{akhmetshin} R. R. Akhmetshin et al.,  
Phys. Lett. B {\bf 489} (2000) 125.
\bibitem{delcourt} B. Delcourt et al.,  
Phys. Lett. B {\bf 113} (1982) 93, Erratum - ibid, B {\bf 115} (1982) 503.
\bibitem{antonelli} A. Antonelli et al.,  Phys. Lett. B {\bf 212} (1988) 133.
\bibitem{artuso} M. Artuso et al.,  Phys. Rev. Lett.  {\bf 62} (1992) 78.
\bibitem{busculic} D. Busculic et al., Z. Phys. C  {\bf 74} (1997) 263.
\bibitem{inami} K. Inami et al., Phys. Lett. B {\bf 672} (2009) 209.
\bibitem{pich} A. Pich,  Phys. Lett. B {\bf 196} (1987) 561.
\bibitem{gilman} F. J. Gilman,  Phys. Rev. D.  {\bf 35} (1987) 3541.
\bibitem{braaten} E. Braaten, R. J. Oakes  and S. M. Tse,  
Phys. Rev. D  {\bf 36} (1987) 2188.
\bibitem{kramer} G. Kramer and W. F. Palmer, Z. Phys. C {\bf 39} (1988) 423.
\bibitem{narison} S. Narison and A. Pich,  Phys. Lett. B  {\bf 304} (1993) 359.
\bibitem{decker} R. Decker and R. Mirkes,  Phys. Rev. D  {\bf 47} (1993) 4012.
\bibitem{li} B. A. Li,  Phys. Rev. D  {\bf 57} (1998) 1790.
\bibitem{PHOKHARA} G. Rodrigo, H. Czy\.z, J. H. K\"{u}hn,  
Eur. Phys. J. C {\bf 24} (2002) 71.
\bibitem{TAUOLA} S. Jadach, Z. W\c{a}s, R. Decker, J. H. K\"{u}hn, 
Comput. Phys. Commun. {\bf 76} (1993) 361.
\bibitem{akhmetshin2} R. R. Akhmetshin et al.,  
Phys.  Lett. B  {\bf 489} (2000) 125.
\bibitem{feldman} G. J. Feldman and R. D. Cousins,  
Phys. Rev. D.  {\bf 57} (1998) 3873.
\bibitem{bergfeld} T. Bergfeld et al., Phys. Rev. Lett.  {\bf 79} (1997) 2406.


\end{thebibliography}
\end{document}